\documentclass[11pt]{article}
\pdfoutput=1
\usepackage{amssymb, graphicx, hyperref}
\graphicspath{{figs/}}
\newcommand{\fig}[2]{\includegraphics[width=#1\textwidth]{#2}}
\newcommand{\offline}{\mbox{$\overline{\rm Off}$\hspace{.05em}\raisebox{.3ex}{$\underline{\rm line}$}}}

\begin{document}
\title{\bfseries Analysis of Fluorescence Telescope Data Using Machine Learning Methods
	}
\author{Mikhail Zotov$^{a,*}$ and Pavel Zakharov$^b$
	for the JEM-EUSO collaboration\\[3mm]
	$^a$ D.V.~Skobeltsyn Institute of Nuclear Physics,\\
	M.V.~Lomonosov Moscow State University, Moscow 119991 Russia\\
	$^*$ e-mail: \texttt{zotov@eas.sinp.msu.ru}\\[2mm]
	$^b$ Faculty of Computational Mathematics and Cybernetics,\\
	M.V. Lomonosov Moscow State University, Moscow 119991 Russia
	}
\date{Presented at the 38th Russian Cosmic Ray Conference, Moscow, 2024}

\maketitle
\begin{abstract}

	Fluorescence telescopes are among the key instruments used for studying ultra-high energy cosmic rays in all modern experiments. We use model data for a small ground-based telescope EUSO-TA to try some methods of machine learning and neural networks for recognizing tracks of extensive air showers in its data and for reconstruction of energy and arrival directions of primary particles. We also comment on the opportunities to use this approach for other fluorescence telescopes and outline possible ways of improving the performance of the suggested methods.

\end{abstract}

\section{Introduction}

Fluorescence telescopes (FTs) are an important part of all major modern
experiments aimed at studying ultra-high energy cosmic rays (UHECRs,
$E\ge1$~EeV), both the Pierre Auger Observatory \cite{b1} and the
Telescope Array \cite{b2}.  FTs register scintillation light emitted
from nitrogen molecules in the air excited during the development of
extensive air showers (EASs) generated by UHECRs. Measurements are
performed in clear moonless nights in the near-UV band. The future
cosmic ray observatories are also planned to employ the fluorescence
technique, both in ground-based experiments like GCOS~\cite{b3} and in
orbital experiments like K-EUSO~\cite{b4} or POEMMA~\cite{b5}.  In
comparison with surface detectors, be it scintillation detectors of
Telescope Array or water tanks of Auger, FTs allow one to estimate
energy of primary UHECRs calorimetrically, decreasing the dependence on
models of hadronic interactions that cannot be verified experimentally
at these energies (see, e.g.,~\cite{b1} for details).

However, both Auger and Telescope Array experiments employ large,
expensive telescopes. Another approach, based on using small and
comparatively cheap FTs is being developed since mid-2010s: these are
the FAST~\cite{b6} and CRAFFT projects~\cite{b7} that develop telescopes
with an aperture around 1~m$^2$. Besides this, a small EUSO-TA telescope
built by the JEM-EUSO collaboration has been in operation at the
Telescope Array site since 2015~\cite{b8}. Using the Telescope Array
trigger, it demonstrated the ability to register cosmic rays with
energies at around 1~EeV and higher.

Besides their main purpose, FTs have proved to be multi-purpose
instruments that can be used to register other fast phenomena that
manifest themselves via UV emission in the atmosphere. In particular,
they are used by the Auger collaboration to study ELVEs~\cite{b9} and by
the Telescope Array to study terrestrial gamma-ray flashes~\cite{b10}.
The first orbital FT, TUS, was also able to register flashes of multiple
different types~\cite{b11}.  Besides this, the joint Russian-Italian
fluorescence telescope Mini-EUSO (``UF atmosfera'' in the Russian Space
Program) has been in operation onboard the International Space Station
since 2019, creating a UV map of the nocturnal atmosphere of the Earth
but also registering meteors, transient luminous events and other
phenomena~\cite{b12, b13}. All these considerations, together with the
current state of the future orbital missions, made us consider the
scientific potential of a small FT of the EUSO-TA type in terms of
recognizing tracks of EASs and the following reconstruction of energy
and arrival directions of primary UHECRs, see also~\cite{b14}.  Using
model data, we developed a few methods based on machine learning and
neural networks to solve the two tasks. In what follows, we describe the
main ideas of the approach, the difficulties we had to overcome, and
outline possible directions of the further improvement of the suggested
methods.

\section{The EUSO-TA Telescope and Simulated Data}

EUSO-TA is a refractor type telescope equipped with two Fresnel lenses
with a diameter of 1 m each, and a concave focal surface (FS) built of
$48\times48$ pixels.  The total field of view (FoV) of the instrument is
$10.5^\circ\times10.5^\circ$, and the time resolution (also called
gate-time unit, GTU) equals 2.5~$\mu$s. Each record consists of 128
data frames of the size of $48\times48$ (``snapshots'' of the focal
surface) written each GTU. The telescope can operate at different
elevation angles. A detailed description of EUSO-TA can be found
in~\cite{b8}.

To create data sets for training and testing our models, we employed
CONEX~\cite{b15} and EUSO-\offline\ codes~\cite{b16}. The first program
was used to simulate EASs produced by proton primaries in the energy
range 5--100~EeV distributed uniformly vs.\ azimuth angles and
proportionally to $\cos\theta$ for zenith angles $0^\circ$--$70^\circ$.
QGSJETII-04~\cite{b17} was chosen as a model of hadronic interactions.
EUSO-\offline\ was used to simulate the detector response.  The shower
cores were put within the projection of the FoV of EUSO-TA on the
ground. The distance between shower cores and the telescope varied from
2~km to 40~km, depending on the energy of primary particles.  Uniform
background illumination had an average rate of 1~photon count per pixel
per GTU, as is typical in real observations during moonless nights.  We
tried data sets with energy distributed quasi-uniformly and
log-uniformly. Neither of the two methods demonstrated superiotity in
terms of the accuracy of energy reconstruction.

Simulated signals used in our models were requested to produce software
triggers. No other quality cuts were applied to the data set, except
omitting signals that had just a few hit pixels in one of the corners of
the telescope FS. In particular, we did not demand that tracks contained
intervals with the shower maximum, even though this can strongly improve
the accuracy of energy reconstruction.

\section{Track Recognition}

The problem of recognizing hit pixels of tracks produced by EASs can be
considered as semantic segmentation, i.e., a computer vision task that
assigns a class label to each pixel of an image. We are only interested
in finding two classes (types) of pixels: those that form a track and
all the rest.  In~\cite{b18}, we presented a solution of this task based
on a convolutional encoder-decoder. Here we show a totally different
approach, based on ``classical'' machine learning methods that do not have
some shortcomings of neural networks~\cite{b19}.

The main and most common method of classical machine learning for
classification and regression problems is gradient boosting. It is based
on reducing the error by constructing an ensemble of simple models (most
often decision trees), each of which at a new step tries to reduce the
error of the previous step~\cite{b20}. Gradient boosting has the
following advantages:

\begin{itemize}
	\item it does not require designing the model architecture;

	\item it does not require data normalization;

	\item it works with categorical factors;

	\item it does not require graphics accelerators for training and inference.
\end{itemize}

In particular, our experiment was run on a basic version of Google Colab
with an Intel Xeon CPU with two vCPUs and 13~GB of RAM.

The track segmentation model on a separate frame is a gradient boosting
model that solves the problem of binary classification. That is, for
each pixel, based on the factors collected for it, it predicts an
estimate of the probability that this particular pixel on the current
frame is a part of the track. Thus, each data record consisting of 128
frames of the size of $48\times48$ generates almost 295 thousand
potential points for training. However, a major part of the frames does
not contain a track and therefore can be excluded from training. In
frames where a track is present, it usually occupies 1--10\% of all
pixels in the frame.

Data for the model was sampled from the original data set as follows:
\begin{enumerate}

	\item The size of the sample (the number of events) for the model was
		chosen.

	\item In each event, only those frames where the proportion of hit pixels
		was greater than 1\% of the total were selected.

	\item The value of photon counts in each pixel was normalized
		(divided) by the maximum value in the frame.

	\item Track labels were modified: pixels with relative normalized
		values less than 0.25 were excluded from the list of hit pixels.

	\item Factors were generated for pixels in each frame.

	\item The pixels were randomly divided into training, validation, and
		testing samples in the ratio of 60/20/20.

\end{enumerate}

\begin{figure}[!ht]

	\fig{1}{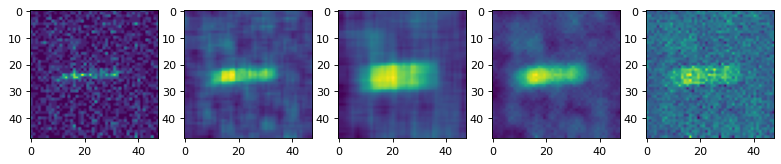}
	\caption{Visual representation of the generated features. From left
	to right: the initial signal with background illumination, 5-neighbor
	average, 10-neighbor average, a bilateral filter, and an inverse
	bilateral filter.}

	\label{fig1}
\end{figure}

The factors for the model are various filters from the OpenCV
library~\cite{b21} that are applied to each frame and transform the
value of the signal in each pixel according to a certain rule. Standard
linear and inverse transformations were selected as filters, the
parameters for which were chosen basing on preliminary training (on 1--2
examples). A visual representation of the filters that were used further
in the model can be seen in Fig.~\ref{fig1}. An in-depth discussion of
image filtering can be found in~\cite{b22}.

After creating the samples, parameters for the boosting algorithm
(XGBoost) were set. Gradient boosting can not build connections between
points on a model level as convolutional neural networks do. Instead,
for each pixel and each GTU it relies on seven numbers. These are a
label assigned to the pixel and six features: four different filterings
as shown in Fig.~\ref{fig1},  the maximum photon count in the data frame
for this particular GTU, and the photon count in the pixel normalized by
the maximum value in the frame. Thus, the model does not need to see the
whole frame at once to process connections between pixels, and therefore
the pixel-level sampling was implemented.

\begin{figure}[!ht]

	\fig{1}{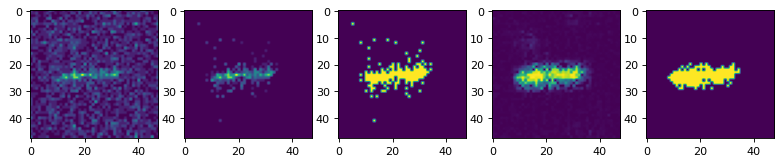}
	\caption{The track recognition model in action. From left to right:
	the initial signal with background illumination, the same signal with
	the background removed, the focal surface with hit pixels (yellow)
	and all the rest, predicted probabilities for the FS pixels to belong
	to the track, recognized hit pixels (yellow).}

	\label{fig2}
\end{figure}

Data from the training and validation samples were fed to the model to
control overfitting. Metrics were calculated on the test data. Next, to
translate the probability estimates into binary labels, the optimal
cutoff was selected: various cutoff values from~0 to~1.0 with a step
of~0.01 were considered and the value equal to~0.67 giving the maximum
F1-score on the test data was determined.  An example of applying the
model to a data frame can be seen in Fig.~\ref{fig2}.  More details
about methods of selecting a cutoff can be found in~\cite{b23}.

To determine the optimal parameters, several versions of the models were
trained, in which the number of iterations and the size of the training
sample were sequentially increased.  Our tests have revealed that a
reasonable balance between the time necessary for training the model and
the values of metrics that characterize its performance is reached for
samples that include just 2000 events. The final model had the following
parameters: the number of trees and the maximum tree depth were equal to
3000 and~3 respectively. The learning rate was chosen to be~0.05.

Two performance metrics that are often used in tasks of image
recognition to evaluate the accuracy of models are the area under the
precision-recall curve (PR AUC) and the balanced accuracy, which is just
the mean of the true positive and true negative rates. Our model
demonstrated PR AUC and the balanced accuracy equal to 0.900 and 0.898
respectively.  These values are slightly lower than 0.958 and 0.949
obtained with the convolutional encoder-decoder, but they needed 16
times less sample size to train the model. This significantly lowers
demands on computing resources and also suggest a way for training
similar models with limited training samples.

\section{Reconstruction of Energy and Arrival Directions}

The conventional procedure of reconstructing an event registered with a
fluorescence telescope begins with the determination of the
shower-detector plane (SDP), see, e.g.,~\cite{b24}. The accuracy of
finding the SDP depends on a number of factors, among them the length of
the track on the FS of the detector. Next, the shower axis is
reconstructed using the timing information from hit pixels. Finally,
after the shower geometry is defined, reconstruction of energy becomes
possible taking into account all contributing light sources
(fluorescence and Cherenkov light, multiple-scattered light) and
necessary corrections for ``invisible energy'' carried away by neutrinos
and high-energy muons.  Still, it should be remarked that the accuracy
of monocular reconstruction is limited even with huge FTs at the Auger
and Telescope Array installations when the measured angular speed does
not change much over the observed track length. This can be the case,
e.g., for short tracks.  That is why the best results are usually
obtained when data of fluorescence telescopes are combined with
information from surface detectors (so called hybrid reconstruction). In
this case, the energy resolution of the Auger FT defined as
event-to-event statistical uncertainty equals 10\%~\cite{b24}.

For EUSO-TA, the task of energy reconstruction is complicated by its
small field of view, so that only a small part of a shower track is
available in a typical record, see a detailed discussion in~\cite{b25}.
Due to its comparatively coarse time resolution (2.5~$\mu$s compared to
100~ns for the Auger and Telescope Array FTs), data records of EUSO-TA
have less accurate information on the temporal development of EAS.  It
should also be remarked that luminosity of a shower signal in a
ground-based FT strongly depends on the distance from the telescope to
the shower axis.  As a result, a shower originated from a strong but
distant EAS can be dimmer than a nearby shower generated by a less
energetic primary.

In 2023, we developed a simple 6-layer convolutional neural network
(CNN)~\cite{b26} for reconstructing energy of events registered by the
fluorescence telescope of the stratospheric EUSO-SPB2
experiment~\cite{b27}.  It demonstrated decent accuracy with the mean
absolute percent error (MAPE) of the order of 10\% when just integrated
tracks were used as input data.  However, the timing information is
crucial for a successful reconstruction of events registered by a
ground-based FT. This applies to estimating both energy and arrival
directions because ``flat'' images of integrated tracks do not allow one
to determine neither real luminosity of an extensive air shower, nor its
azimuth with respect to the telescope.

This suggested arranging input data in the form of a ``stack'' of images
of the focal surface.  Each image represented a ``screenshot'' of the FS
at one particular GTU. Thus, a sequence of images made at consecutive
moments of time provided information about the temporal development of
events.  The simulated data had trigger GTUs near the beginning of
records. Our tests demonstrated that in most cases it is sufficient to
utilize data of the first 12 GTUs, thus strongly reducing demands on the
amount of memory needed for model training.

The initial CNN developed for EUSO-SPB2 was optimized to use more
complex data representation necessary for EUSO-TA.  The data set used
for training the neural network consisted of 52 thousand events with
20\% of them acting as a validation set.  All data records were linearly
scaled so that the brightest signal was equal to~1.  As a result, we
were able to teach the CNN to reconstruct both energy and arrival
directions of the simulated events.  The loss function, which is needed
at the stage of training, was chosen according to the reconstructed
parameters. The MAPE was employed when energy was the only parameter to
be reconstructed.  In other cases, we used either mean squared error or
mean absolute error.  Angular separation complemented one of the two
functions in case we were only interested in reconstructing arrival
directions. It is interesting to mention that, contrary to the
conventional approach, we do not need to determine the shower geometry
before reconstructing the energy of a primary particle. The CNN is able
to reconstruct energy independently of the arrival direction, and vice
versa, as well as it is able to solve both tasks simultaneously.

\begin{figure}[!ht]

	\centering
	\fig{.8}{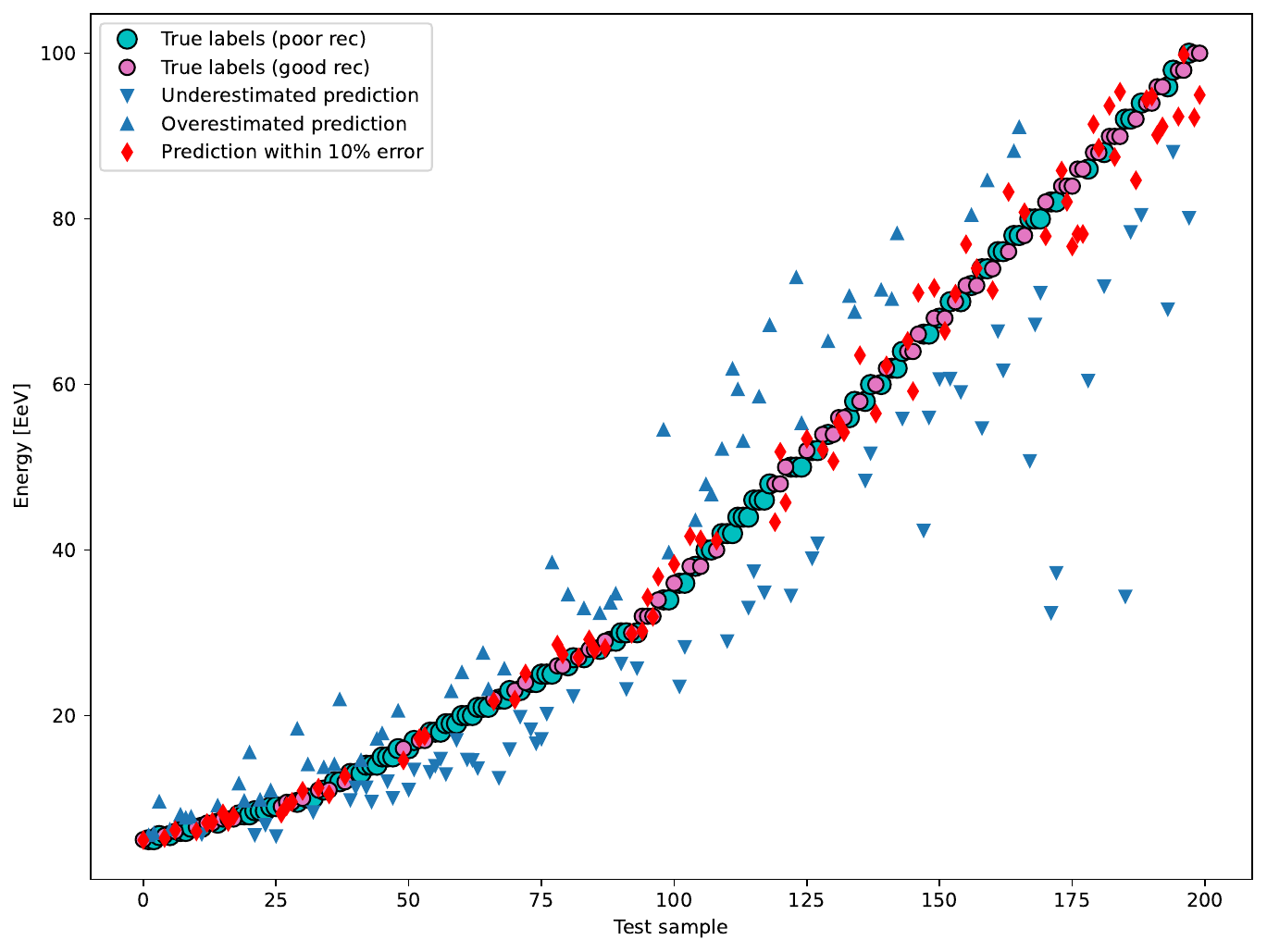}
	\caption{Example of energy reconstruction for EUSO-TA on a test
	sample consisting of 200 events with energies in the range 5--100~EeV.
	Circles: true (simulated) energies of primary protons, diamonds:
	predicted energies with the absolute percentage error $<10\%$,
	triangles: other predictions. MAPE equals 17.7\% in this test.}

	\label{fig3}

\end{figure}

The trained models were tested on data sets that included 200--500 events
from the whole range of energies and arrival directions.  In a typical
test, energy was reconstructed with MAPE in the range 15--20\%, see an
example in Fig.~\ref{fig3}. The mean error was around 1\% with the standard
deviation $\sim20\%$. The largest errors as expressed in MAPE were
observed for events at the lower end of energies and arriving from
directions nearly orthogonal to the axis of the field of view of EUSO-TA
due to the signal being confined to mere 1-2 GTUs of the respective
record.

\begin{figure}[!ht]

	\centering
	\fig{.8}{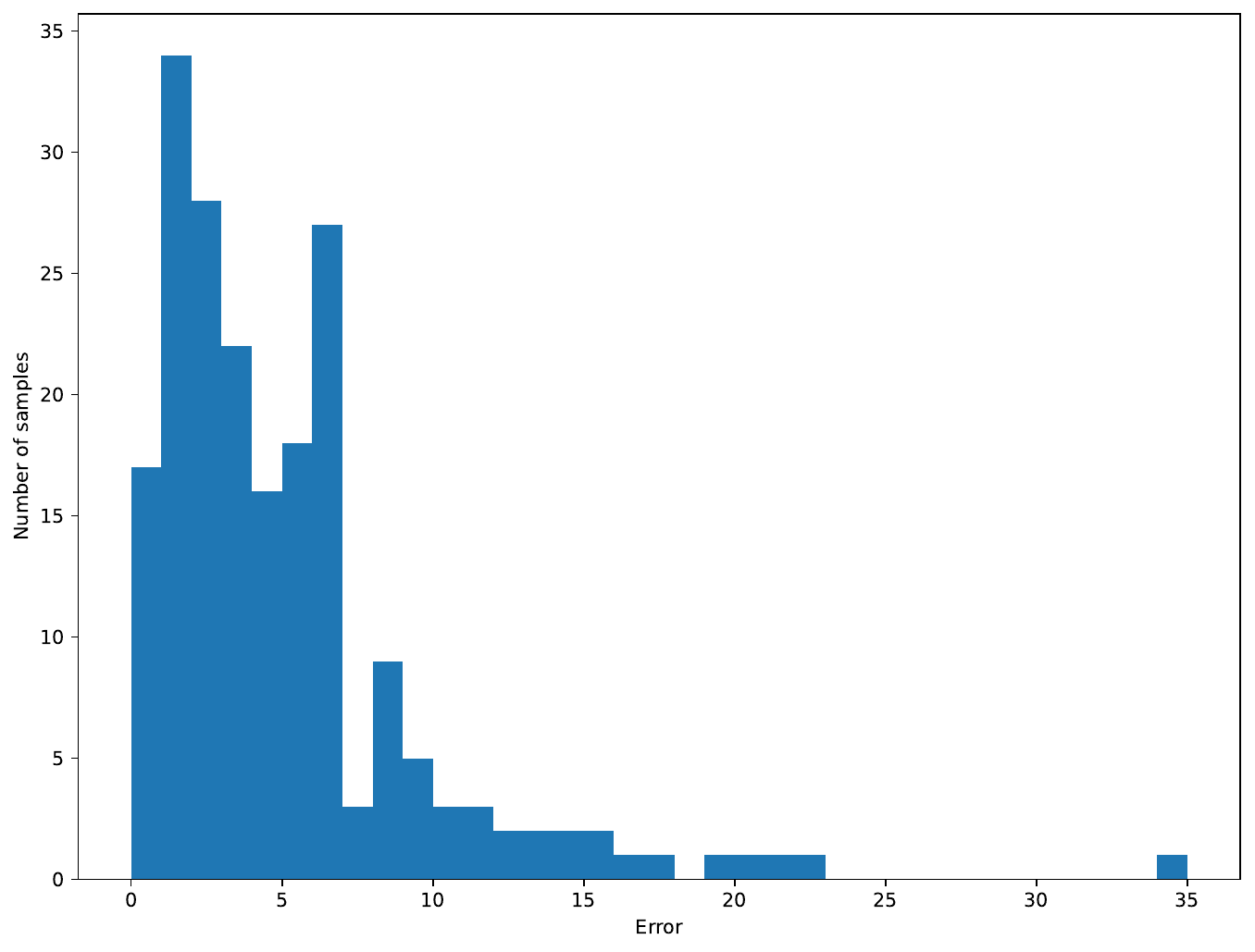}
	\caption{Angular separation between true (simulated) and
	reconstructed arrival directions for the same sample as in Fig.~3.
	Errors are expressed in degrees. The median error equals $3.9^\circ$.}

	\label{fig4}

\end{figure}

Zenith and azimuth angles were reconstructed with mean errors
$\le0.5^\circ$ and standard deviations of the order of
$4^\circ$--$5^\circ$ and $15^\circ$ respectively. The median value of
angular separation between true and reconstructed arrival directions was
typically around $4^\circ$--$5^\circ$.  Expectedly, the largest errors
took place for events with azimuth angles nearly orthogonal to the axis
of the FoV and for events touching only the very corners of the focal
surface. Azimuth angles that were along the axis of the telescope
$\pm20^\circ$ were reconstructed with the least errors\footnote{It is
worth mentioning that EASs from these directions produce triggers more
often than those arriving from other directions. As a result,
along-the-axis showers were better presented in the training sample than
the others.}.  Fig.~\ref{fig4} presents a typical distribution of angular
separation between true and reconstructed arrival directions.

\section{Discussion and Conclusions}

We have developed a proof-of-concept convolutional neural network, which
is able to reconstruct energy and arrival directions of UHECRs for a
small ground-based fluorescence telescope EUSO-TA using simulated data.
The same CNN is able to solve this task for the FT pointed in nadir from
a stratospheric instrument EUSO-SPB2.  The accuracy of reconstruction
cannot be directly compared with the much more sophisticated and large
FTs working at Pierre Auger Observatory and the Telescope Array, but we
think the first results are promising and can be further improved. In
particular, models can be trained on a virtual focal surface which is
much larger than the real one to teach the neural network to reconstruct
missing information from tracks that otherwise hit just a few pixels.
More obvious ways of improving the results are to increase the size of
the training data set, to tune the hyperparameters of the CNN, to
perform training on subsets of the energy range and arrival directions,
etc.

We have also developed two methods of recognizing EAS tracks in the
focal surface of fluorescence telescopes of both EUSO-TA and EUSO-SPB2.
One of them employs gradient boosting, one of the well-known methods of
classical machine learning. The second one uses a convolutional
encoder-decoder, i.e., an artificial neural network with a special
architecture. Both methods demonstrated accuracy $\gtrsim0.9$ with the
neural network performing slightly better. However, the method based on
gradient boosting needed a considerably smaller training set and put
lower demands on computing resources.

We believe that the suggested methods for EAS track recognition and for
reconstruction of energy and arrival directions of primary ultra-high
energy cosmic rays are generic and can be applied to other fluorescence
telescopes though the architecture of the neural networks will probably
need certain modifications to take into account features of a particular
instrument.

\bigskip
\bigskip
\textbf{Funding:}
M.Z.\ is supported by grant 22-62-00010 from the Russian Science Foundation.


\begin{thebibliography}{30}

	\bibitem{b1} The Pierre Auger Collaboration,  Nuclear Instruments and
		Methods in Physics Research A 798, 172 (2015).
		\href{https://doi.org/10.1016/j.nima.2015.06.058}{https://doi.org/10.1016/j.nima.2015.06.058}

	\bibitem{b2}  T. Abu-Zayyad, R. Aida, M. Allen et al., Nuclear
		Instruments and Methods in Physics Research A 689, 87 (2012).
		\href{https://doi.org/10.1016/j.nima.2012.05.079}{https://doi.org/10.1016/j.nima.2012.05.079}

	\bibitem{b3} J.R. H{\"o}randel, in Proc. 37th International Cosmic Ray
		Conference, Berlin, Germany, PoS 395, 27 (2021).
		\href{https://doi.org/10.22323/1.395.0027}{https://doi.org/10.22323/1.395.0027}

	\bibitem{b4} P. Klimov, M. Battisti, A. Belov et al., Universe 8, 88
		(2022). \href{https://doi.org/10.3390/universe8020088}{https://doi.org/10.3390/universe8020088}

	\bibitem{b5} A.V. Olinto, J. Krizmanic, J.H. Adams et al., Journal of
		Cosmology and Astroparticle Physics 2021, 007 (2021).
		\href{https://doi.org/10.1088/1475-7516/2021/06/007}{https://doi.org/10.1088/1475-7516/2021/06/007}

	\bibitem{b6} M. Malacari, J. Farmer, T. Fujii et al., Astroparticle
		Physics 119, 102430 (2020).
		\href{https://doi.org/10.1016/j.astropartphys.2020.102430}{https://doi.org/10.1016/j.astropartphys.2020.102430}

	\bibitem{b7} Y. Tameda, T. Tomida, M. Yamamoto et al., Progress of
		Theoretical and Experimental Physics 2019, 043F01 (2019).
		\href{https://doi.org/10.1093/ptep/ptz025}{https://doi.org/10.1093/ptep/ptz025}

	\bibitem{b8} J.H. Adams, S. Ahmad, J.-N. Albert et al., Experimental
		Astronomy 40, 301 (2015).
		\href{https://doi.org/10.1007/s10686-015-9441-6}{https://doi.org/10.1007/s10686-015-9441-6}

	\bibitem{b9} A. Tonachini for the Pierre Auger Collaboration, Proc.
		33rd International Cosmic Ray Conference, Rio de Janeiro, Brazil,
		0676 (2013).

	\bibitem{b10} J. W. Belz, P. R. Krehbiel, J. Remington et al., JGR
		Atmospheres 125, e2019JD031940 (2020).
		\href{https://doi.org/10.1029/2019JD031940}{https://doi.org/10.1029/2019JD031940}

	\bibitem{b11} P. Klimov, S. Sharakin, M. Zotov, M.E. Bertaina and F.
		Fenu, Proc. 37th International Cosmic Ray Conference, Berlin,
		Germany, PoS 395, 316 (2021).
		\href{https://doi.org/10.22323/1.395.0316}{https://doi.org/10.22323/1.395.0316}

	\bibitem{b12} S. Bacholle, P. Barrillon, M. Battisti at al.,
		Astrophysical Journal Supplement Series 253, 36 (2021).
		\href{https://doi.org/10.3847/1538-4365/abd93d}{https://doi.org/10.3847/1538-4365/abd93d}

	\bibitem{b13} D. Barghini, M. Battisti, A. Belov et al., Astronomy \&
		Astrophysics 687,  A304 (2024).
		\href{https://doi.org/10.1051/0004-6361/202449236}{https://doi.org/10.1051/0004-6361/202449236}

	\bibitem{b14} P. Klimov, A. Belov, M. Zotov, S. Sharakin, D. Chernov,
		``Concept of a small fluorescence telescope for registering
		extensive air showers,'' \href{https://events.sinp.msu.ru/event/12/attachments/665/1299/01_07_%D0%9F%D0%9A%D0%9B1_%D0%9A%D0%BB%D0%B8%D0%BC%D0%BE%D0%B2.pdf}{report}
		at the 38th Russian Cosmic Ray
		Conference (2024).

	\bibitem{b15} T. Bergmann, R. Engel, D. Heck, N. Kalmykov, S.
		Ostapchenko, T. Pierog, T. Thouw, and K. Werner, Astroparticle
		Physics 26, 420 (2007).
		\href{http://dx.doi.org/10.1016/j.astropartphys.2006.08.005}{http://dx.doi.org/10.1016/j.astropartphys.2006.08.005}

	\bibitem{b16} S. Abe, J.H. Adams, D. Allard et al., Journal of
		Instrumentation 19, P01007 (2024).
		\href{https://dx.doi.org/10.1088/1748-0221/19/01/P01007}{https://dx.doi.org/10.1088/1748-0221/19/01/P01007}

	\bibitem{b17} S. Ostapchenko, Nuclear Physics B Proceedings
		Supplements 151, 143 (2006).
		\href{https://doi.org/10.1016/j.nuclphysbps.2005.07.026}{https://doi.org/10.1016/j.nuclphysbps.2005.07.026}

	\bibitem{b18} M. Zotov,
		\href{https://doi.org/10.48550/arXiv.2408.02440}{arXiv:2408.02440}
		(2024).

	\bibitem{b19} N. O' Mahony, S. Campbell, A. Carvalho, S.
		Harapanahalli, G. Velasco-Hernandez, L. Krpalkova, D. Riordan and
		J. Walsh, in Advances in Computer Vision Proceedings of the 2019
		Computer Vision Conference (CVC). Springer Nature Switzerland AG,
		pp.\ 128-144 (2020).
		\href{https://doi.org/10.1007/978-3-030-17795-9}{https://doi.org/10.1007/978-3-030-17795-9}

	\bibitem{b20} A. Natekin and A. Knoll, Frontiers in Neurorobotics 7,
		21 (2013).
		\href{https://doi.org/10.3389/fnbot.2013.00021}{https://doi.org/10.3389/fnbot.2013.00021}

	\bibitem{b21} G. Bradski, Dr. Dobb's Journal of Software Tools,
		2236121 (2000). \href{https://opencv.org}{https://opencv.org}

	\bibitem{b22} B. Desai, M. Paliwal, K.K. Nagwanshi,
		\href{https://doi.org/10.48550/arXiv.2207.06481}{arXiv:2207.06481}
		(2022).

	\bibitem{b23} L. Ferrer,
		\href{https://doi.org/10.48550/arXiv.2209.05355}{arXiv:2209.05355}
		(2022).

	\bibitem{b24} J. Abraham, P. Abreu, M. Aglietta et al., Nuclear
		Instruments and Methods in Physics Research A 620, 227 (2010).
		\href{https://doi.org/10.1016/j.nima.2010.04.023}{https://doi.org/10.1016/j.nima.2010.04.023}

	\bibitem{b25} J.H. Adams, L. Anchordoqui, D. Barghini et al.
		Astroparticle Physics 163, 103007 (2024).
		\href{https://doi.org/10.1016/j.astropartphys.2024.103007}{https://doi.org/10.1016/j.astropartphys.2024.103007}

	\bibitem{b26} G. Filippatos and M. Zotov, in Proc. 38th International
		Cosmic Ray Conference, Nagoya, Japan, PoS 444, 234 (2023).
		\href{https://doi.org/10.22323/1.444.0234}{https://doi.org/10.22323/1.444.0234}

	\bibitem{b27} J.H. Adams, D. Allard, P. Alldredge et al.,
		Astroparticle Physics 165, 103046 (2024).
		\href{https://doi.org/10.1016/j.astropartphys.2024.103046}{https://doi.org/10.1016/j.astropartphys.2024.103046}

\end{thebibliography}
\end{document}